\documentclass{jfm}
\usepackage{graphicx,bm}
\def\dt{{\rm d}t}
\def\d{{\rm d}}
\def\dk{{\rm d}k}

\def\x{{\bm x}}
\def\k{{\bm k}}
\def\etal{{\it et al. }}
\def\norm#1{\left|\mkern-2mu\left|#1\right|\mkern-2mu\right|}

\title[Impeded inverse transfer in the CHM model of quasi-geostrophic 
flows]{Impeded inverse energy transfer in the Charney--Hasegawa--Mima 
model of quasi-geostrophic flows}
\author[C.V. Tran and D. G. Dritschel]
{C\ls H\ls U\ls O\ls N\ls G\ns V.\ns T\ls R\ls A\ls N \ns \and
D\ls A\ls V\ls I\ls D\ns G.\ns D\ls R\ls I\ls T\ls S\ls C\ls H\ls E\ls L}
\affiliation{School of Mathematics and Statistics, University of St Andrews, 
St Andrews KY16 9SS, UK}

\begin{document}

\maketitle

\begin{abstract}

The behaviour of turbulent flows within the single-layer quasi-geostrophic
(Charney--Hasegawa--Mima) model is shown to be strongly dependent on the
Rossby deformation wavenumber $\lambda$ (or free-surface elasticity). 
Herein, we derive a bound on the inverse energy transfer, specifically on 
the growth rate $\d\ell/\dt$ of the characteristic length scale $\ell$ 
representing the energy centroid. It is found that 
$\d\ell/\dt\le2\norm q_\infty/(\ell_s\lambda^2)$, where $\norm q_\infty$ 
is the supremum of the potential vorticity and $\ell_s$ represents the 
potential enstrophy centroid of the reservoir, both invariant. This result 
implies that in the potential energy dominated regime 
($\ell\ge\ell_s\gg\lambda^{-1}$), the inverse energy transfer is strongly 
impeded, in the sense that under the usual time scale no significant 
transfer of energy to larger scales occurs. The physical implication is 
that the elasticity of the free surface impedes turbulent energy transfer 
in wavenumber space, effectively rendering large-scale vortices long-lived 
and inactive. Results from numerical simulations of forced-dissipative 
turbulence confirm this prediction.

\end{abstract}

\section{Introduction}

The atmosphere and oceans are hugely influenced by the background planetary 
rotation and stable density stratification. These features virtually suppress 
motion in the vertical direction, rendering geophysical fluid systems, 
intrinsically vast in their horizontal extent compared with their limited 
depth, approximately two-dimensional (2D) at large scales. Many models 
derived from the theory of quasi-geostrophy exploit this simplification.
One such model, governed by the Charney--Hasegawa--Mima (CHM) equation, 
is the subject of the present study.

The CHM equation, which describes the evolution of potential vorticity in
a shallow fluid layer with a free surface, takes the form (Pedlosky 1987)
\begin{eqnarray}
\label{CHM}
\frac{\partial q}{\partial t}+J(\psi,q) &=& 0.
\end{eqnarray}
Here $\psi(x,y,t)$, the streamfunction, is assumed to be periodic in both
$x$ and $y$. Note that $\psi$ is proportional to the free surface height
anomaly. The quantity $q=(\Delta-\lambda^2)\psi$ is the potential 
vorticity,\footnote{This form of potential vorticity ignores the 
so-called $\beta$-term. For some studies concerning the effects of 
this term, see Kukhakin \& Orszag (1996), Okunu \& Masuda (2003) and 
Smith (2004).} where $\lambda$ is the Rossby deformation wavenumber 
($\lambda=f/c$, where $f$ is the Coriolis frequency and $c$ is the 
small-scale gravity waves speed, both assumed constant). The differential 
operators $J(\cdot,\cdot)$ and $\Delta$ are, respectively, the Jacobian 
and two-dimensional Laplacian. Eq.~(\ref{CHM}) expresses material 
conservation of potential vorticity. It also governs the evolution of 
quasi-2D fluctuations of the electrostatic potential in a plane 
perpendicular to a strong magnetic field applied uniformly to a plasma, 
in which case $\psi$ is the electrostatic potential and $\lambda^{-1}$ 
is the ion Larmor radius (Hasegawa \& Mima 1978). 

The CHM equation has been a subject of active research for decades (see 
for example, Hasegawa \& Mima 1978; Fyfe \& Montgomery 1979; Yanase \& 
Yamada 1984; Larichev \& McWilliams 1991; Ottaviani \& Krommes 1992; 
Kukharkin, Orszag \& Yakhot 1995; Iwayama, Shepherd \& Watanabe 2002; 
Arbic \& Flierl 2003; Tran \& Bowman 2003). A majority of these studies 
have been concerned with the dual transfer of the total energy 
$\langle-\psi q\rangle/2=\langle|\nabla\psi|^2\rangle/2+\lambda^2\langle
\psi^2\rangle/2$ and potential enstrophy $\langle\Delta\psi q\rangle/2=
\langle|\Delta\psi|^2\rangle/2+\lambda^2\langle|\nabla\psi|^2\rangle/2$, 
which are conserved by the CHM nonlinearities. Here $\langle\cdot\rangle$ 
denotes a spatial average. The kinetic energy and the usual enstrophy are 
$\langle|\nabla\psi|^2\rangle/2$ and $\langle|\Delta\psi|^2\rangle/2$, 
respectively (these are not conserved). The quantity 
$\lambda^2\langle\psi^2\rangle/2$ is the potential energy. In addition
to the total energy and potential enstrophy, the CHM dynamics also 
conserve an infinite class of potential vorticity norms, including
the $L^\infty$ norm $\norm q_\infty$. Similar to 2D Navier--Stokes (NS) 
turbulence, the dual conservation law implies that the total energy 
(potential enstrophy) is preferentially transferred to lower (higher) 
wavenumbers. Unlike 2D NS turbulence, each of the invariants in CHM 
turbulence consists of two components. The presence of the second 
component is due to the elasticity of the free surface, which is known 
to have a fundamental influence on the dynamics. A remarkable effect is 
that the inverse energy transfer for wavenumbers $k\ll\lambda$ (the 
so-called potential energy regime)\footnote{For scales smaller than 
$\lambda^{-1}=c/f$, the motion is inherently three-dimensional (cf. 
Dritschel \& de la Torre Ju\'arez 1996; Dritschel, de la Torre Ju\'arez 
\& Ambaum 1999) and therefore poorly described by (\ref{CHM}).} is 
predominantly a transfer of the potential energy (Tran \& Bowman 2003), 
i.e. of the quadratic quantity $\langle\psi^2\rangle$. 

Numerical studies of CHM turbulence appear to suggest that in the 
potential energy regime, the inverse cascade is strongly impeded. For 
turbulence decaying from an initial energy reservoir, \cite{Larichev91} 
observe a ``slowness'' in the turbulent evolution. Iwayama \etal (2002) 
notice the formation of a sharp spectral peak, a manifestation of an 
inverse energy transfer with limited extent in wavenumber space. 
\cite{Arbic03} find that the inverse cascade slows down and that vortex 
merging takes a very long time. For forced turbulence, Kukharkin \etal 
(1995) observe the formation of vortical ``quasicrystals'', states of 
long-lived inactive vortices. While these observations are consistent 
with the finding of Tran \& Bowman (2003) that the inverse energy transfer 
involves virtually no kinetic energy when $k\ll\lambda$, they are not 
fully explained by this fact. A more satisfactory explanation would be 
that the inverse transfer of potential energy is also impeded in this regime. 

In this Note we derive an upper bound for the growth rate of a 
characteristic length scale associated with the inverse energy transfer.
We infer from this upper bound that in the potential energy regime,
the inverse energy transfer is strongly impeded by the elasticity of
the free surface. An interesting physical interpretation of this effect 
is that this elasticity prohibits the merger of large-scale vortices to 
form larger energy-carrying scales, effectively rendering such vortices 
long-lived and inactive. Although this effect is deduced in the framework 
of unforced and inviscid dynamics, it also manifests itself in a 
forced-dissipative version of (\ref{CHM}), as is illustrated by some 
numerical results. 

\section{Inverse energy transfer}

The nonlinear transfer of energy to lower wavenumbers of an energy reservoir
(when it occurs) necessarily entails an increase of a suitably defined length 
scale representing the energy centroid of the reservoir. Here we consider the 
length scale $\ell$ defined by  
\begin{eqnarray}
\label{ell}
\ell=\frac{\langle(-\Delta)^{-1/2}\psi q\rangle}{\langle\psi q\rangle}
=\frac{\langle|(-\Delta)^{1/4}\psi|^2\rangle+\lambda^2\langle|(-\Delta)^
{-1/4}\psi|^2\rangle}{\langle|\nabla\psi|^2\rangle+\lambda^2\langle
\psi^2\rangle}. 
\end{eqnarray}
In (\ref{ell}) the operator $(-\Delta)^\theta$ is defined by
$(-\Delta)^\theta\widehat\psi(\k)=k^{2\theta}\widehat\psi(\k)$, where
$k=|\k|$ is the wavenumber and $\widehat\psi(\k)$ the Fourier transform
of $\psi(\x)$. The growth rate $\d\ell/\dt$ provides a quantitative 
measure of the inverse transfer. In particular, a smaller rate 
corresponds to an inverse transfer with a more limited extent in 
wavenumber space than one with a greater rate. In this respect, 
$\d\ell/\dt$ may be considered as a measure of locality of the 
inverse transfer, in the sense that for a given energy distribution 
a smaller rate implies a more local transfer than a greater one.

In the case where most of the kinetic energy resides in wavenumbers 
$k\ll\lambda$ (the potential energy regime),  
$\langle|(-\Delta)^{1/4}\psi|^2\rangle$ is negligible compared with 
$\lambda^2\langle|(-\Delta)^{-1/4}\psi|^2\rangle$, and similarly 
$\langle|\nabla\psi|^2\rangle$ is negligible compared with
$\lambda^2\langle\psi^2\rangle$. More precisely, in this regime we 
have $$\frac{\langle|(-\Delta)^{1/4}\psi|^2\rangle}{\lambda^2\langle
|(-\Delta)^{-1/4}\psi|^2\rangle}\le\frac{\langle|\nabla\psi|^2\rangle}
{\lambda^2\langle\psi^2\rangle}\ll1,$$ where the first inequality is due
to H\"older's inequality (see also Tran \& Shepherd 2002 and Tran 2004), 
implying 
$\ell\approx\langle|(-\Delta)^{-1/4}\psi|^2\rangle/\langle\psi^2\rangle$. 
For an energy reservoir in the potential energy regime, the length scale 
$\ell$ specifies where, in wavenumber space, most of $\langle\psi^2\rangle$ 
(or equivalently most of the potential energy) is distributed. Its growth 
rate is thus the rate at which the spectral profile of the potential 
energy proceeds toward lower wavenumbers.

The evolution equation for the characteristic length scale $\ell$, defined 
in the preceding paragraphs, is obtained by multiplying (\ref{CHM}) by 
$(-\Delta)^{-1/2}\psi/\langle\psi q\rangle$ and taking the spatial average 
of the resulting equation:
\begin{eqnarray}
\label{evol1}
\frac{1}{2}\frac{\d\ell}{\dt} &=&
\frac{\langle(-\Delta)^{-1/2}\psi J(\psi,q)\rangle}
{\langle|\nabla\psi|^2\rangle+\lambda^2\langle\psi^2\rangle}
=-\frac{\langle q J(\psi,(-\Delta)^{-1/2}\psi)\rangle}
{\langle|\nabla\psi|^2\rangle+\lambda^2\langle\psi^2\rangle}\nonumber\\
&\le&
\frac{\langle|q||\nabla\psi||\nabla(-\Delta)^{-1/2}\psi|\rangle}
{\langle|\nabla\psi|^2\rangle+\lambda^2\langle\psi^2\rangle}
\le \frac{\langle|\nabla\psi|^2\rangle^{1/2}\langle\psi^2\rangle^{1/2}}
{\langle|\nabla\psi|^2\rangle+\lambda^2\langle\psi^2\rangle}
\norm q_\infty,
\end{eqnarray} 
where the various manipulations of the nonlinear term are straightforward.
In particular, the last step can be readily seen by expressing the 
streamfunction in terms of a Fourier series.

The ratio in front of $\norm q_\infty$ on the right-hand side of 
(\ref{evol1}) can be closely estimated in terms of conserved quantities. 
For that purpose, let us denote by $\ell_s$ the conserved length scale 
representing the enstrophy centroid of the reservoir:
\begin{eqnarray}
\label{enscentroid}
\ell_s &=& 
\frac{\langle-\psi q\rangle^{1/2}}{\langle\Delta\psi q\rangle^{1/2}}=\frac
{\left(\langle|\nabla\psi|^2\rangle+\lambda^2\langle\psi^2\rangle\right)^{1/2}}
{\left(\langle|\Delta\psi|^2\rangle+\lambda^2\langle|\nabla\psi|^2\rangle
\right)^{1/2}}.
\end{eqnarray} 
Solving (\ref{enscentroid}) for $\langle\psi^2\rangle$ and 
substituting the result into the said ratio we obtain
\begin{eqnarray}
\label{ratio}
\frac{\langle|\nabla\psi|^2\rangle^{1/2}\langle\psi^2\rangle^{1/2}}
{\langle|\nabla\psi|^2\rangle+\lambda^2\langle\psi^2\rangle} 
&=& \frac{\langle|\nabla\psi|^2\rangle^{1/2}
\left[\ell_s^2\langle|\Delta\psi|^2\rangle+(\ell_s^2\lambda^2-1)\langle
|\nabla\psi|^2\rangle\right]^{1/2}}{\lambda\left[\langle|\nabla\psi|^2
\rangle+\ell_s^2\langle|\Delta\psi|^2\rangle+(\ell_s^2\lambda^2-1)\langle
|\nabla\psi|^2\rangle\right]}\nonumber\\ 
&\le& \frac{\langle|\nabla\psi|^2\rangle^{1/2}
\left(\langle|\Delta\psi|^2\rangle+\lambda^2\langle
|\nabla\psi|^2\rangle\right)^{1/2}}{\ell_s\lambda\left(\langle|
\Delta\psi|^2\rangle+\lambda^2\langle
|\nabla\psi|^2\rangle\right)}\nonumber\\ 
&=& \frac{\langle|\nabla\psi|^2\rangle^{1/2}}{\ell_s\lambda
\left(\langle|\Delta\psi|^2\rangle+\lambda^2\langle
|\nabla\psi|^2\rangle\right)^{1/2}}
\le\frac{1}{\ell_s\lambda^2}.
\end{eqnarray} 
This is a very sharp estimate, especially for a reservoir in the 
potential energy regime. We note that in general the inequality 
$\ell\ge\ell_s$ holds. This is a consequence of the H\"older inequality
and can be shown as follows. Let us denote by $U(k)$ the power spectrum 
of $\langle-(-\Delta)^{-1/2}\psi q\rangle$. The length sales $\ell$
and $\ell_s$ can then be expressed in terms of $U(k)$ as 
$$\ell=\frac{\int U(k)\,\dk}{\int kU(k)\,\dk}$$ and 
$$\ell_s=\left(\frac{\int kU(k)\,\dk}{\int k^3U(k)\,\dk}\right)^{1/2}.$$ 
Now by H\"older's inequality we have 
$$\int kU(k)\,\dk\le\left(\int U(k)\,\dk\right)^{2/3}
\left(\int k^3U(k)\,\dk\right)^{1/3},$$ from which the inequality
$\ell\ge\ell_s$ indeed follows immediately.

Putting these results together we have
\begin{eqnarray}
\label{evol2}
\frac{\d\ell}{\dt} &\le& 2\frac{\norm q_\infty}{\ell_s\lambda^2}
\le 2\frac{\norm q_\infty}{\ell_s^2\lambda^2}\ell
\end{eqnarray}
using $\ell\ge\ell_s$. Hence the exponential growth rate of $\ell$ is 
bounded from above by $2\norm q_\infty/(\ell_s^2\lambda^2)$. Given a 
finite $\norm q_\infty$, the upper bound $2\norm q_\infty/(\ell_s^2\lambda^2)$ 
can be made arbitrarily small provided $\ell_s^2\lambda^2$ is sufficiently 
large. This condition is satisfied for an initial energy reservoir in a 
sufficiently remote wavenumber region of the potential energy regime,
where both $\ell_s\lambda\gg1$ and $\ell\lambda\gg1$. Under these 
circumstances, no significant growth of $\ell$ occurs when the initial 
energy reservoir freely spreads out, and hence no significant inverse 
energy transfer is possible. In passing it is worth mentioning that the 
upper bound for the exponential growth rate of $\ell$ in (\ref{evol2}) 
consists of the constant $\lambda$ and the two invariants $\norm q_\infty$ 
and $\ell_s$, and therefore can be completely determined by initial data.

The result in the preceding paragraph is an analytic version of the 
qualitative statement in the literature that the turbulence evolves 
on a rescaled (slower) time (cf. Larichev \& McWilliams 1991; Kukharkin, 
Orszag \& Yakhot 1995). To clarify this point let us rewrite 
(\ref{evol2}) as
\begin{eqnarray}
\label{evol3}
\ell_s^2\lambda^2\frac{\d\ell}{\dt} &\le& 2\norm q_\infty\ell.
\end{eqnarray}
The left-hand side of (\ref{evol3}) is the (instantaneous) growth rate 
of $\ell$ on the slow time $\tau=t/(\ell_s^2\lambda^2)$. For an initial
reservoir with a finite $\norm q_\infty$, the inverse transfer of 
potential energy can be considerable only on this slow time scale. In 
the limit $\ell_s\lambda\rightarrow\infty$, the invariants become 
$\langle\psi^2\rangle/2$ and $\langle|\nabla\psi|^2\rangle/2$ and the 
length scales $\ell$ and $\ell_s$ become
$\langle|(-\Delta)^{-1/4}\psi|^2\rangle/\langle\psi^2\rangle$ and 
$\langle\psi^2\rangle^{1/2}/\langle|\nabla\psi|^2\rangle^{1/2}$, 
respectively. The rescaled time $\tau$ becomes infinitely long. 

An estimate which may be significantly sharper than (\ref{evol2}) can be 
derived by replacing the potential vorticity $q$ in (\ref{evol1}) by the 
relative vorticity $\Delta\psi$. By doing that, instead of (\ref{evol2}), 
we obtain
\begin{eqnarray}
\label{evol4}
\frac{\d\ell}{\dt} &\le& 2\frac{\norm{\Delta\psi}_\infty}
{\ell_s^2\lambda^2}\ell.
\end{eqnarray}
In remote regions of the potential energy regime, where 
$\ell\lambda\ge\ell_s\lambda\gg1$, we expect $\lambda^2\psi$ to 
dominate $\Delta\psi$. Therefore in these regions it is plausible that
$\norm q_\infty\gg\norm{\Delta\psi}_\infty$, allowing (\ref{evol4}) to 
become significantly sharper than (\ref{evol2}). The price for this 
improvement is that $\norm{\Delta\psi}_\infty$ is not conserved. Note
that in such regions $\ell$ and $\ell_s$ are the length scales of 
potential and kinetic energy, respectively.

A straightforward physical interpretation of the above result is that 
the elasticity of the free surface suppresses energy transfer for scales 
larger than the Rossby deformation radius. This effect has a profound 
influence on the vortex dynamics. Let us consider a hypothetical flow 
consisting of vortices (having smooth potential vorticity---no small 
length scales) of comparable scales, which are much larger than the 
deformation radius $\lambda^{-1}$. For these scales the inverse energy 
transfer is strongly impeded, allowing for virtually no formation of 
larger energy-carrying scales. This keeps the vortices from merging, 
effectively establishing a state of long-lived inactive vortices. This 
may explain the crystallisation of vortices observed by Kukharkin \etal 
(1995). The impeded inverse energy transfer implies a similar effect on 
the direct transfer of potential enstrophy, a consequence of the dual 
conservation law. One can see from the ``symmetry'' of the dual 
conservation law that the available potential enstrophy for direct 
transfer depends on the inversely-transferred energy. Similarly, the 
available energy for inverse transfer depends on the directly-transferred 
potential enstrophy. Hence impeded inverse energy transfer means impeded 
direct potential enstrophy transfer (and vice versa). This implies that 
smooth potential vorticity with scales $\gg\lambda^{-1}$ remains smooth.  

\section{Numerical results}

We now consider results obtained from numerical simulations of a 
forced-dissipative version of (\ref{CHM}). These convincingly illustrate 
the strong reduction of the inverse transfer in the potential energy 
regime, and may help explain an observation by Iwayama \etal (2002) in 
their numerical simulations of freely decaying CHM turbulence wherein an 
initial energy reservoir evolves into a single sharp peak. A plausible 
explanation for this observation is that the sharp peak is due to both 
the viscous dissipation (which suppresses the formation of small scales) 
and the impeded inverse transfer deduced in the preceding section: their 
collective effect ``squeezes'' energy into a narrow band of wavenumbers, 
resulting in the observed peak. In our forced-dissipative case we observe 
that in the potential energy regime the inverse energy flux is arrested 
and deposits most of the potential energy at the lower-wavenumber end of 
the quasi-steady energy range. We also observe a relaxation of the kinetic 
energy spectrum of the inertial range (from $k^{-5/3}$ toward $k^{-1}$) 
as the inverse cascade proceeds from $k>\lambda$ to $k<\lambda$. 
 
We simulate Eq. (\ref{CHM}), with the addition of molecular viscosity 
and a body force, in a doubly periodic square $2\pi\times2\pi$ with 
a dealiased $1024^2$ pseudospectral method. No large-scale dissipation
mechanisms are employed and no steady dynamics are sought since we are
primarily interested in the near arrest of the inverse energy cascade in
a quasi-steady state. (For an analysis of steady CHM turbulence in the
presence of large-scale damping, see Smith \etal 2002.) Three different 
values of $\lambda$ are used: $\lambda=20,40$ and $60$. The forcing is 
spectrally localised in the wavenumber interval $K=[49.5,50.5]$, in the 
sense that its Fourier components $\widehat f(\k)$ are nonzero only for 
those modes $\k$ having magnitudes $k$ lying in the interval $K$: 
\begin{eqnarray}
\label{forcing}
\widehat{f}(\bm k)&=&\frac{\epsilon}{N}\frac{\widehat{\psi}(\bm{k})}
{\displaystyle\sum_{|\bm{p}|=k}|\widehat{\psi}(\bm{p})|^2}.
\end{eqnarray}
Here $\epsilon=1$ is the constant energy injection rate and~$N$ is the 
number of distinct wavenumbers in~$K$. The viscosity coefficient is 
$\nu=5\times10^{-4}$. The simulations were initialised with the kinetic 
energy spectrum $E(k)=10^{-5}\pi k/(50^2+k^2)$.

\begin{figure}
\centerline{\includegraphics{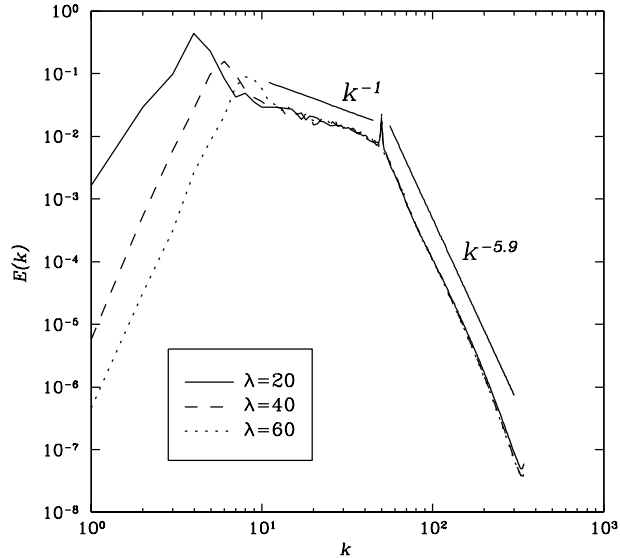}}
\caption{The kinetic energy spectrum $E(k)$ {\it vs.}\ $k$ averaged over 
periods $[139,140]$, $[155,156]$ and $[160,161]$ for $\lambda=20$, $40$ 
and $60$, respectively.} 
\label{atmosp234}
\end{figure} 

Figure \ref{atmosp234} shows the quasi-steady kinetic energy spectrum 
$E(k)$ vs. $k$ 
averaged over periods $[139,140]$, $[155,156]$ and $[160,161]$ for 
$\lambda=20$, $40$ and $60$, respectively. In all three cases, the 
accumulation of energy at low wavenumbers is clearly visible. This is 
attributed to the impeding effect discussed above, and not finite-size 
effects since the energy levels at the lowest wavenumbers are still 
small compared with the energy peaks. This is true for both kinetic 
and potential components. For $\lambda=20$ we have a limited kinetic 
energy regime in the wavenumber interval $[20,50]$. Before the energy 
peak reaches $\lambda=20$, the dynamics of this kinetic energy regime 
are expected to resemble those of 2D NS turbulence: the energy cascades 
to low wavenumbers via a $k^{-5/3}$ kinetic energy spectrum. We indeed 
observe a discernible $k^{-5/3}$ spectrum (not shown) in the wavenumber 
range $[20,50]$ before the energy peak reaches $\lambda=20$. The dynamics 
after the energy peak passes $\lambda$ undergo profound changes. Most 
notably, the growth of potential energy occurs mainly around the energy 
peak, consistent with the reduction of the inverse energy transfer in 
the conservative case. Another remarkable change is in the spectral 
slope of the energy spectrum from $k=\lambda=20$ up to the forcing 
wavenumber $k=50$. The slope of the kinetic energy spectrum in this 
region appears to relax from $-5/3$ to $-1$. The quasi-steady\footnote
{Note that after the energy peak passes $\lambda$, growth of energy 
is due primarily to that of the potential energy and the kinetic 
energy becomes quasi-steady.} spectrum becomes $k^{-1}$ with an 
excessive accumulation of energy at the lower-wavenumber end. This 
unexpected feature is common to the other two cases, for which hardly 
any inverse transfer region exists between the forcing wavenumber and 
$\lambda$ and for which no clear $k^{-5/3}$ spectrum is formed initially.

\begin{figure}
    \begin{center}
    \includegraphics[scale=0.625]{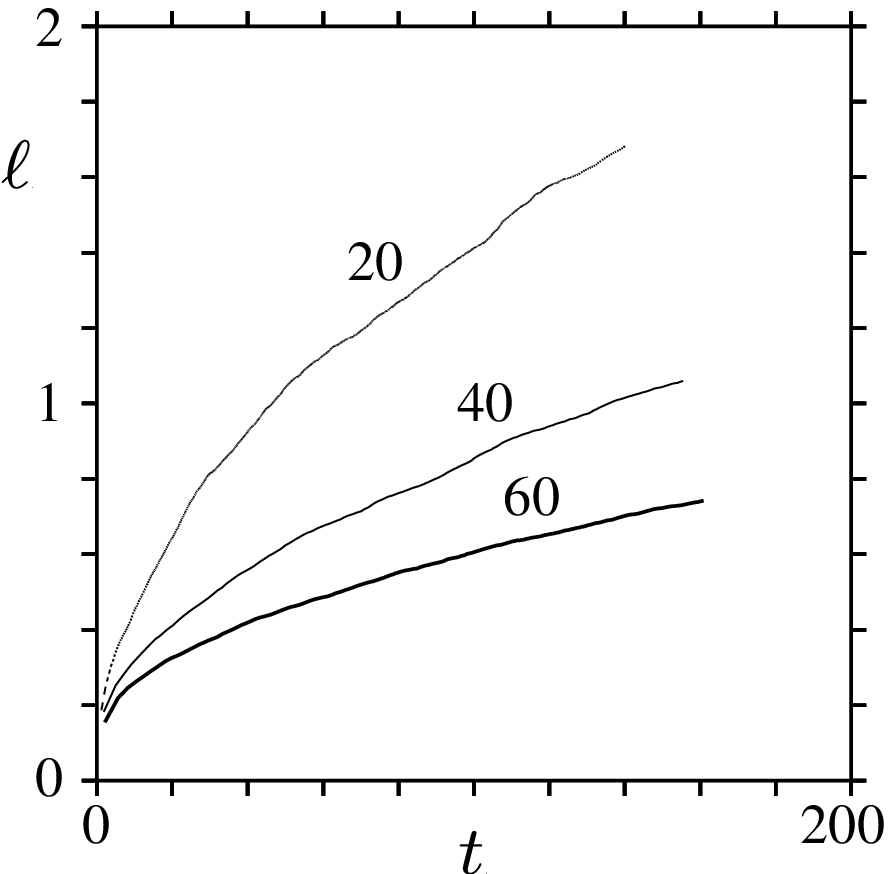} \,\,
    \includegraphics[scale=0.625]{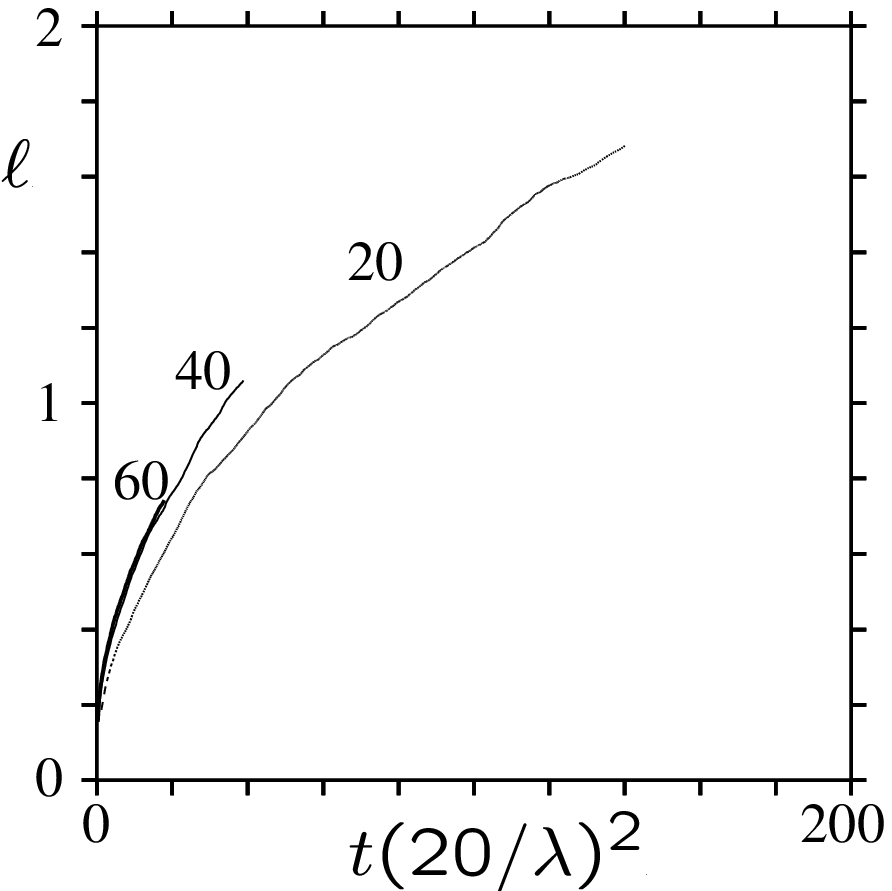} 
    \end{center}
\caption{The energy centroid $\ell$ versus time $t$ (left) and scaled
time $t(20/\lambda)^2$ (right) for $\lambda=20$, $40$ 
and $60$, as labelled.} 
\label{ell_vs_time}
\end{figure} 

The evolution of the energy centroid $\ell$ for these three cases is
shown in Figure \ref{ell_vs_time}, on the left versus unscaled time $t$
and on the right versus scaled time $t(20/\lambda)^2$.  Both the 
reduction in ${\d\ell}/{\dt}$ with increasing $\lambda$ {\it and} 
the collapse of the curves when plotted versus scaled time are 
consistent with the growth rate bound (\ref{evol2}).\footnote
{We cannot directly compare the numerical growth rates with 
the theoretical bound because the imposed forcing violates 
conservation of $\norm{q}_\infty$ and $\ell_s$ used 
in (\ref{evol2}).}
Only the $\lambda = 20$ case stands out slightly, due to 
an early kinetic energy transfer which is virtually absent in 
the other two cases.  

\section{Summary and discussion}

The main result of this work is an upper bound for the growth rate of
the characteristic length scale associated with the inverse energy 
transfer in CHM turbulence. This upper bound is expressible in terms of 
the supremum of the potential vorticity, the characteristic length scale
representing the potential enstrophy centroid of the reservoir and the 
Rossby deformation radius. These are constants of motion, allowing the 
bound to be completely described in terms of initial data. It is found 
that in remote regions of the potential energy regime, the turbulent 
transfer from an initial energy reservoir becomes strongly impeded, in 
the sense that hardly any potential energy is transferred to lower 
wavenumbers. This effect has been illustrated by numerical simulations 
of a forced-dissipative version of the CHM equation. The physical 
implication of the present finding is that the elasticity of the free
surface hinders the nonlinear transfer of potential energy to larger scales. 

An interesting implication of the present result is that vortices having
smooth potential vorticity and scales much larger than the Rossby 
deformation radius are kept from merging to form larger energy-carrying 
scales and remain smooth. Such vortices are inactive and long-lived 
compared to their counterparts in 2D NS turbulence.

The present theory is fully consistent with other results for both forced
(Kukharkin \etal 1995) and freely decaying (Larichev \& McWilliams 1991; 
Iwayama \etal 2002; Arbic \& Flierl 2003) CHM turbulence, as discussed above.
These, we argue, are explained by the growth rate bound (\ref{evol3}) and 
its physical interpretations given at the end of \S\,2. 

Analyses of atmospheric and oceanic data yield results that could plausibly 
be explained by the present work. \cite{Boer83} find that the spectral flux 
in atmospheric wind data is arrested with essentially no inertial range. 
\cite{Scott05} find the inverse energy transfer in the oceans to be 
qualitatively similar to the result of \cite{Boer83}. These are in 
remarkable agreement with the idealised numerical results noted above and 
with the present findings.

Iwayama \etal (2002) observe ``well-behaved'' tails of the potential 
vorticity probability density function (PDF)---much steeper than observed
in NS turbulence, see their figure 5---and raise the question why it 
is that CHM turbulence in the potential energy regime exhibits no 
intermittency. A speculative answer provided by the authors is based 
on an asymptotic version of the CHM equation wherein the relative 
vorticity is dropped from the time-derivative term: 
$$\lambda^2\,\frac{\partial\psi}{\partial t}+J(\Delta\psi,\psi) = 0.$$ 
This equation, as noted by Iwayama \etal (2002), describes advection of
$\psi$ by $\Delta\psi$ (i.e. by the smaller scales of motion). These 
authors argue that this may lead to more ``random walk'' behaviour, 
effectively suppressing intermittency. The present study provides an 
alternative answer: the reduction of turbulent transfer both renders 
large-scale vortices inactive and discourages erratic small-scale 
motion. This also can account for the observed steep PDF. In some sense 
CHM turbulence in the potential energy regime is far less turbulent 
than its 2D NS counterpart. 

\begin{acknowledgments}
The numerical simulations in \S\,3 were carried out, using a computer 
code developed by John Bowman, when CVT was a Pacific Institute for the 
Mathematical Sciences postdoctoral fellow at the University of Alberta. 
Constructive comments from two anonymous referees were very much appreciated.
\end{acknowledgments}

\end{document}